\documentstyle[psfig]{l-aa}

\begin{document}

\thesaurus {06 (04.03.1, 08.02.5)}

\title{UBV(RI)$_{\rm C}$ photometric comparison sequences for \\ symbiotic stars. II}

\author{
       Arne Henden\inst{1}
\and   Ulisse Munari\inst{2}
       }
\offprints{U.Munari}

\institute {
Universities Space Research Association/U. S. Naval Observatory
Flagstaff Station, P. O. Box 1149, Flagstaff AZ 86002-1149, USA
\and
Osservatorio Astronomico di Padova, Sede di Asiago, 
I-36012 Asiago (VI), Italy
}
\date{Received date..............; accepted date................}

\maketitle

\markboth{A.Henden and U.Munari: UBV(RI)$_{\rm C}$ photometric comparison sequences for 
symbiotic stars. II}{A.Henden and U.Munari: UBV(RI)$_{\rm C}$ photometric comparison 
sequences for symbiotic stars. II}

\begin{abstract}
We present accurate UBV(RI)$_{\rm C}$ photometric sequences for an additional
20 symbiotic stars. As for the 20 targets of Paper I, the sequences extend
over wide brightness and color ranges, and are suited to cover both
quiescence and outburst phases. The sequences are intended to assist both
present time photometry as well as measurement of photographic plates from
historical archives.
\end{abstract}
\maketitle

\section{Introduction}

In Paper I (Henden and Munari 2000) we discussed the need for
extended, accurate and homogeneous photometric sequences around symbiotic stars
and how their lack has contributed to our currently poor photometric knowledge
of these interacting binaries. We also reviewed the basic types of
variability for symbiotic stars and their causes, and provided
$UBV(RI)_{\rm C}$ sequences for a first sample of 20 objects.

This Paper II presents accurate photometric comparison sequences for an
additional 20 symbiotic stars using observing strategies, reduction
methodologies and presentation layouts strictly similar to those of Paper I
(to which the reader is referred for all details). The sequences are
intended to allow a general observer to capture on a single CCD frame or to
have in the same eyepiece field of view when inspecting archival
photographic plates: ($a$) enough stars to cover the whole range of known or
expected variability for the given symbiotic star, ($b$) stars of enough
different colors to be able to calibrate the instrumental color equations
and therefore reduce to the standard $UBV(RI)_{\rm C}$ system the collected
data, and ($c$) stars well separated from surrounding ones to avoid blending
at all but the shortest telescope focal lengths. As for Paper I, all
observations have been made with the 1.0-m Ritchey-Chr\'etien telescope of
the U. S. Naval Observatory, Flagstaff Station with a Tektronix/SITe
1024x1024 thinned, backside--illuminated CCD and Johnson {\sl UBV} and
Kron--Cousins {\sl RI} filters.

The sequences around the 20 symbiotic stars of Paper~I have already been
used by Munari et al. (2001) and Jurdana and Munari (2001) to investigate
their long term photometric behavior on the rich collection of Asiago
historical photographic plates. With other observatories with historical
plate archive joining the effort, in a close future the community could have
assembled complete, century-long lightcurves for a significant fraction of
the symbiotic stars. Given their complex, non-periodic and usually slow
photometric evolution with major outbursts appearing erratically and lasting
for years or decades, the understanding of the symbiotic stars and their
role in the general astrophysical context cannot progress much without a
detailed knowledge of their photometric histories and habits.

\begin{table*}
\caption[]{List of program symbiotic stars. The coordinates for the symbiotic stars
are from our observations (equinox J2000.0, mean epoch 2000.3).  The
$e_\alpha$ and $e_\delta$ columns list the errors in milliarcsec for right
ascension and declination, respectively.  $l$ and $b$ are the galactic coordinates. 
Columns 11 and 12 list the coordinates of the field centers in Figures~1 (zoomed-in 5.16$\times$5.16 arcmin
fields), or in Figure~2 when only the larger ones (11$\times$11 arcmin) are
available. The last column indicates which one of the {\sl narrow} (i.e
5.16$\times$5.16 arcmin) or {\sl wide} (e.g. 11.4$\times$11.4 arcmin) fields are given
for the given program star in Figures~1 and 2.}
\centerline{\psfig{file=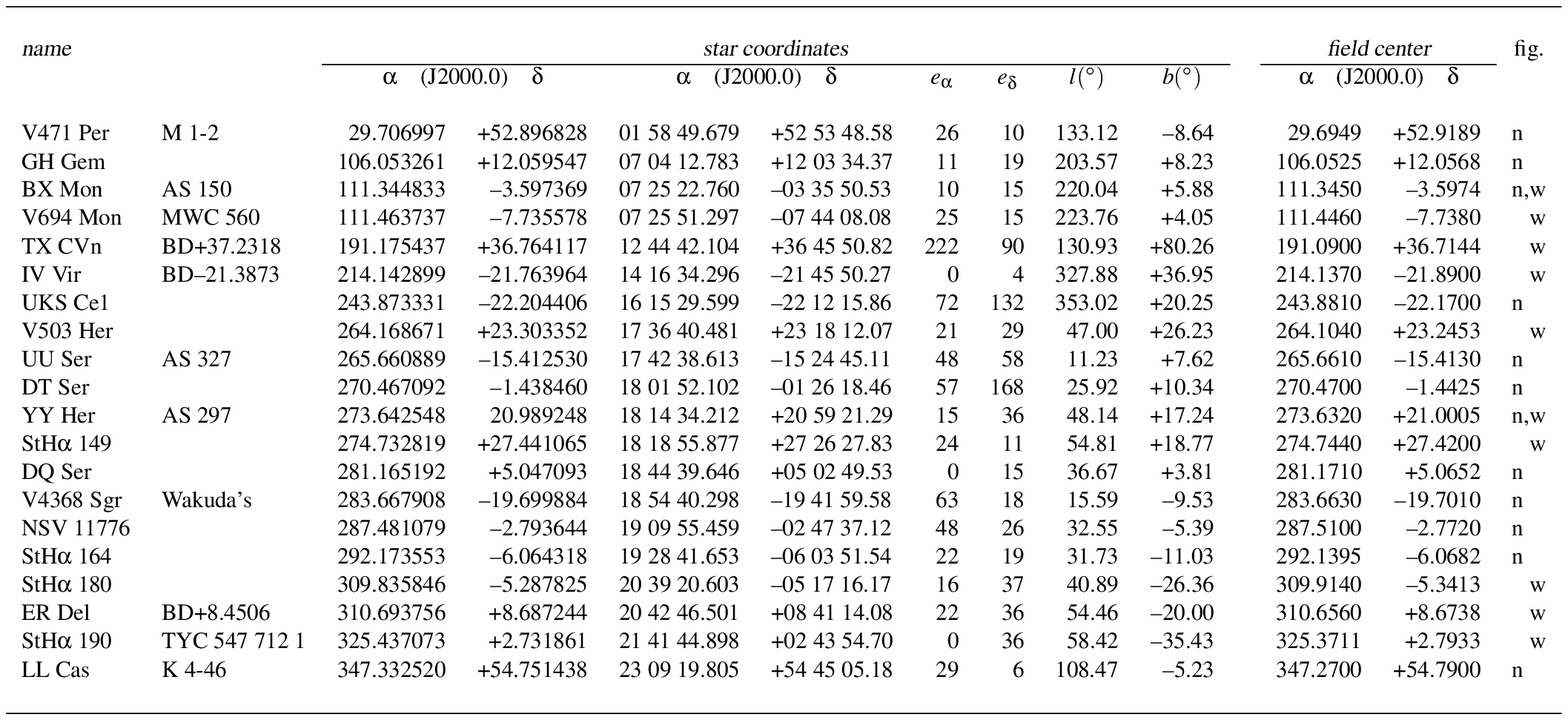,width=18cm}}
\end{table*}

\section{The photometric sequences}

The program symbiotic stars are listed in Table~1.  An average of 11 stars
around each program star have been selected to form the comparison
sequences, which are given in Table~2. The stars included in the sequences
have been chosen and ordered on the basis of the $B$ band magnitude, because
the latter is the band closest to the $m_{pg}$ of the historical
photographic observations and is the better suited to investigate the
variability of symbiotic stars (cf. Kenyon 1986 and Paper~I). The stars
included in the comparison sequences have been checked on at least three
different nights for variability (cf. column $N$ of Table~2).

For 10 objects (V471 Per, GH Gem, UKS Ce1, UU Ser, DT Ser, DQ Ser,
V4368 Sgr, NSV 11776, StH$\alpha$ 164 and LL Cas) the symbiotic star and the
comparison sequence both lie inside a 6.0$\times$6.0 arcmin field (which
match in dimension the Allen 1984 finding charts) and are plotted in
Figure~1. For 8 brighter objects (V694~Mon, TX~CVn, IV~Vir, V503~Her,
ER~Del, StH$\alpha$~149, StH$\alpha$~180 and StH$\alpha$~190) the comparison
sequences are distributed over a 11.4$\times$11.4 arcmin area and are
presented in Figure~2. The remaining two program stars (BX~Mon and YY~Her)
are hybrid cases having their fainter comparison stars identified in
Figure~1 and the brighter, more distant ones in Figure~2.

Finally, for three objects (TX~CVn, ER~Del and StH$\alpha$~190) there are no
stars in the explored 11.4$\times$11.4 arcmin fields brighter than the
symbiotic star itself (particularly important for estimates on photographic
plates). For these program stars we have selected from the Hipparcos/Tycho Catalogue
suitable, nearby and non-variable stars to complement
our comparison sequences. The selected Hipparcos/Tycho stars are listed in Table~3,
where their $B_T, V_T$ magnitudes have been transformed to the standard
Johnson's {\sl UBV} system following the transformation equations provided
in the explanatory notes of the Hipparcos/Tycho Catalogue:
\begin{eqnarray}
    V_J & = & V_T \ - \ 0.090 \times (B-T)_T \\
(B-V)_J & = & 0.850 \times (B-V)_T 
\end{eqnarray}
For sake of completeness we have also reported the $U$, $R_{\rm C}$ and
$I_{\rm C}$ values as estimated from the observed $B-V$ color using the
extensively calibrated relations of Caldwell et al. (1993). These
transformation relations between colors in the $UBV(RI)_{\rm C}$ system gives
accurate results {\sl provided} that the stars belong to the solar
neighborhood population, the reddening is not large and the luminosity
class is roughly known. We have assumed all the selected Tycho objects to be
nearby main sequence stars. Thus, the $U$, $R_{\rm C}$ and $I_{\rm C}$
magnitudes reported in the last three columns of Table~3 may be considered
as guidelines useful for estimating photographic plates.

\begin{table*}
\caption[]{
The comparison sequences around the 20 program symbiotic stars. Positions
for J2000 equinox and a mean epoch 2000.3 are given (errors in arcsec are
derived from different exposures in different bands), together with
magnitudes and colors (errors in magnitudes). The stars in each sequence are
ordered according to fainter $B$ magnitudes. $N$ is the number of observing
nights. The sequences are listed in the same order as the symbiotic stars in Table~1. 
For BX~Mon and YY~Her, the comparison stars lying outside the field of view of
Figure~1 and plotted in Figure~2 are given at the bottom of the sequence,
separated by an empty line.}
\centerline{\psfig{file=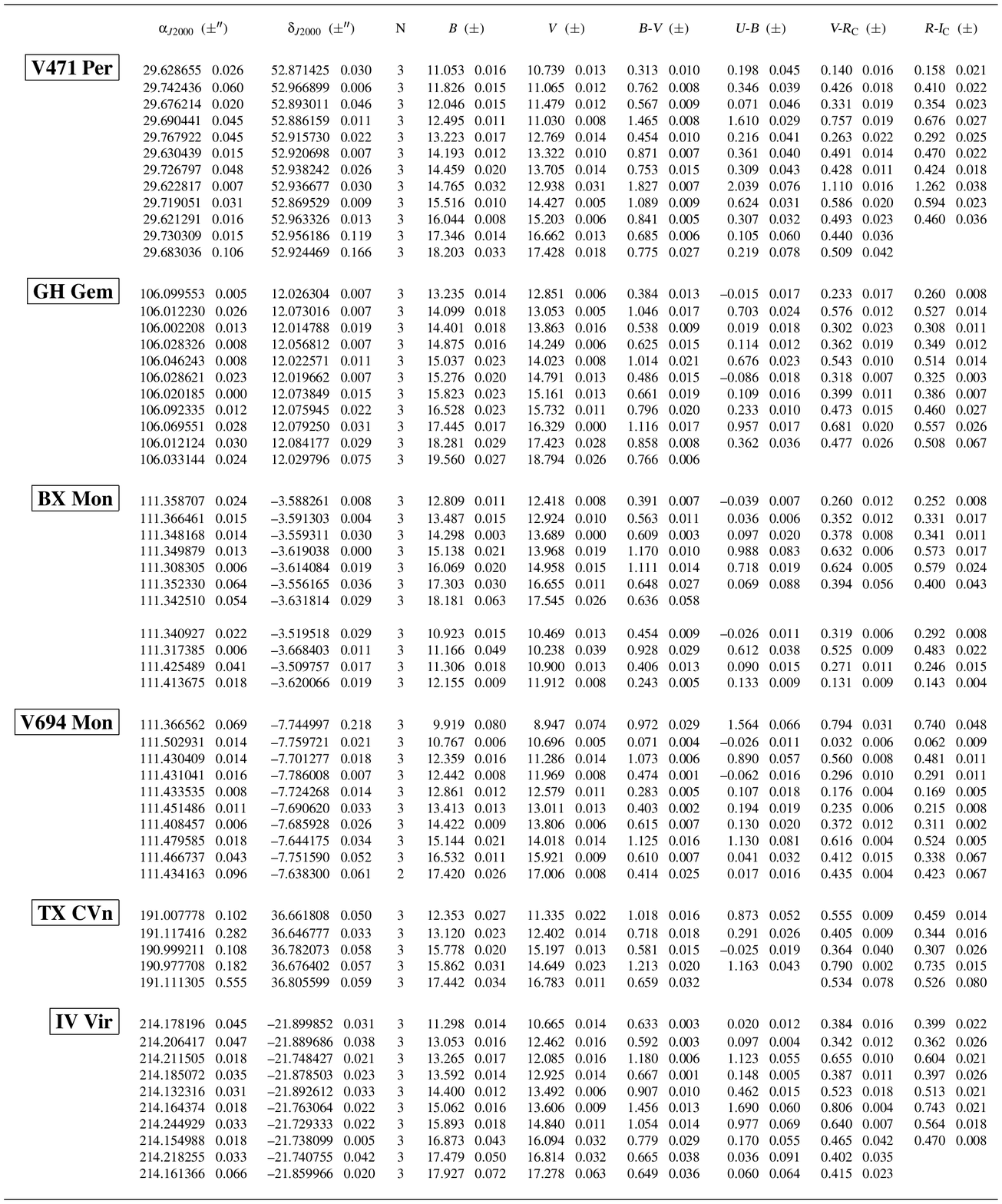,width=18cm}}
\end{table*}

\setcounter{table}{1}
\begin{table*}
\caption[]{({\sl continues})}
\centerline{\psfig{file=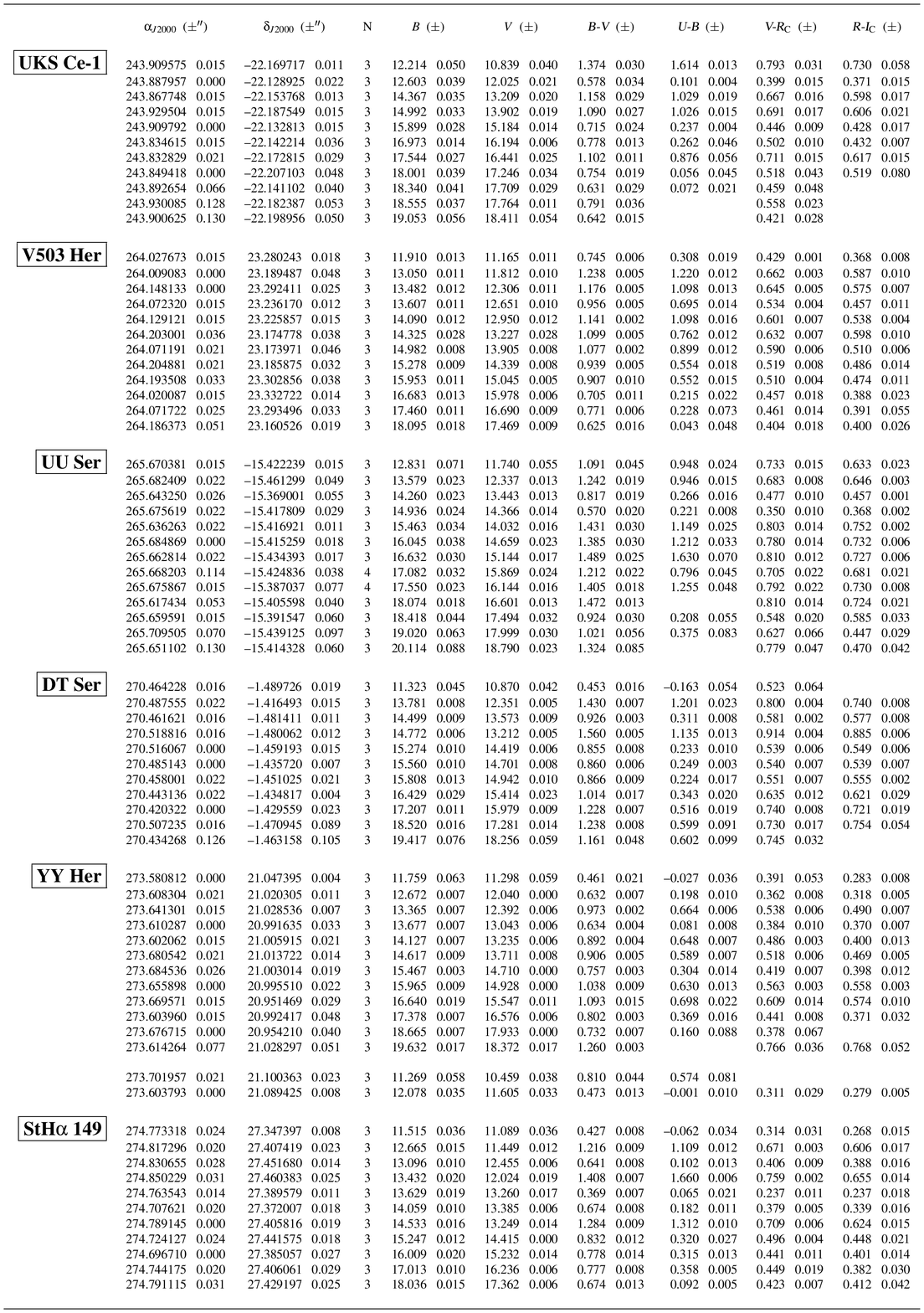,height=23.8cm}}
\end{table*}

\setcounter{table}{1}
\begin{table*}
\caption[]{({\sl continues})}
\centerline{\psfig{file=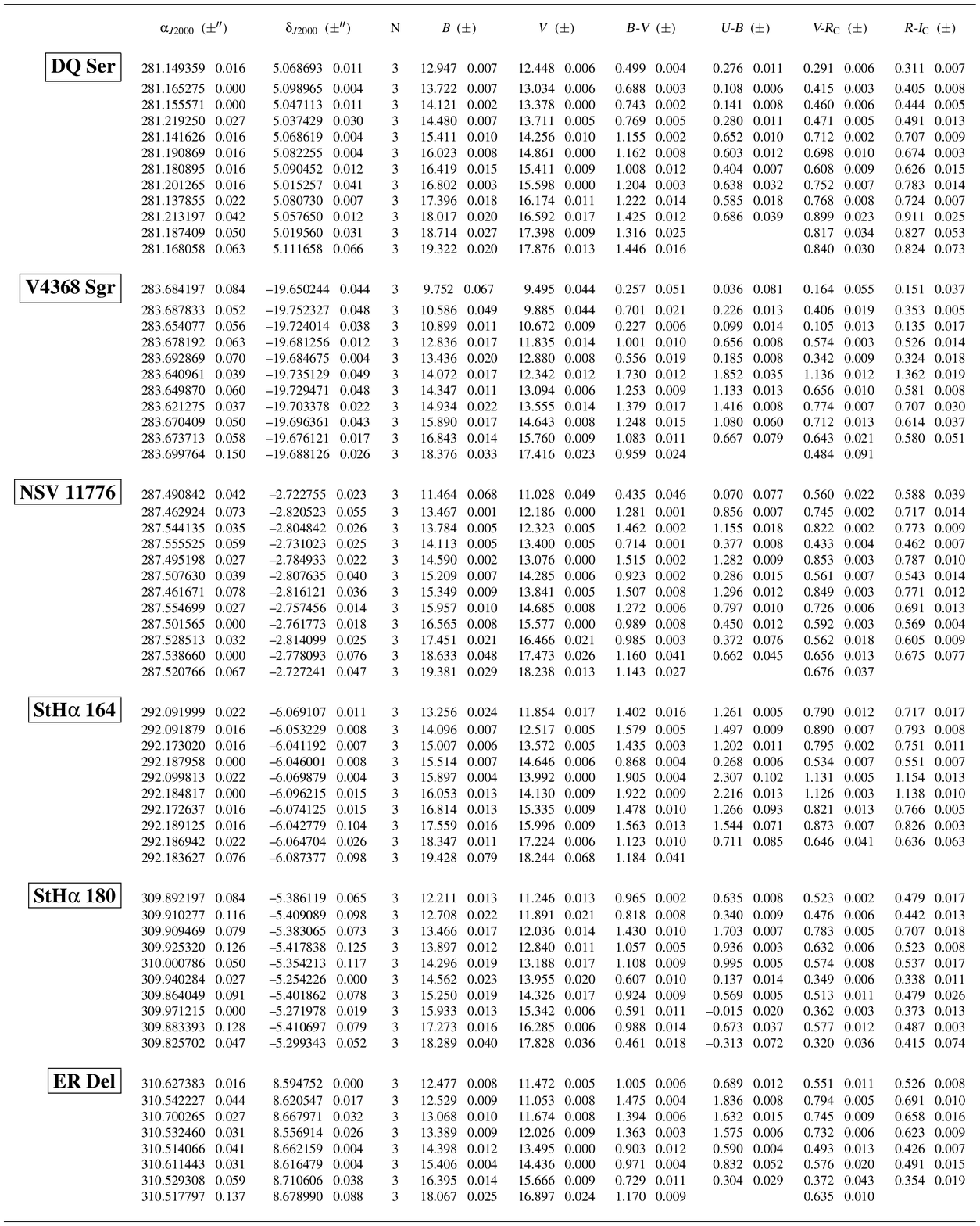,height=23.8cm}}
\end{table*}

\setcounter{table}{1}
\begin{table*}[ht!]
\caption[]{({\sl continues})}
\centerline{\psfig{file=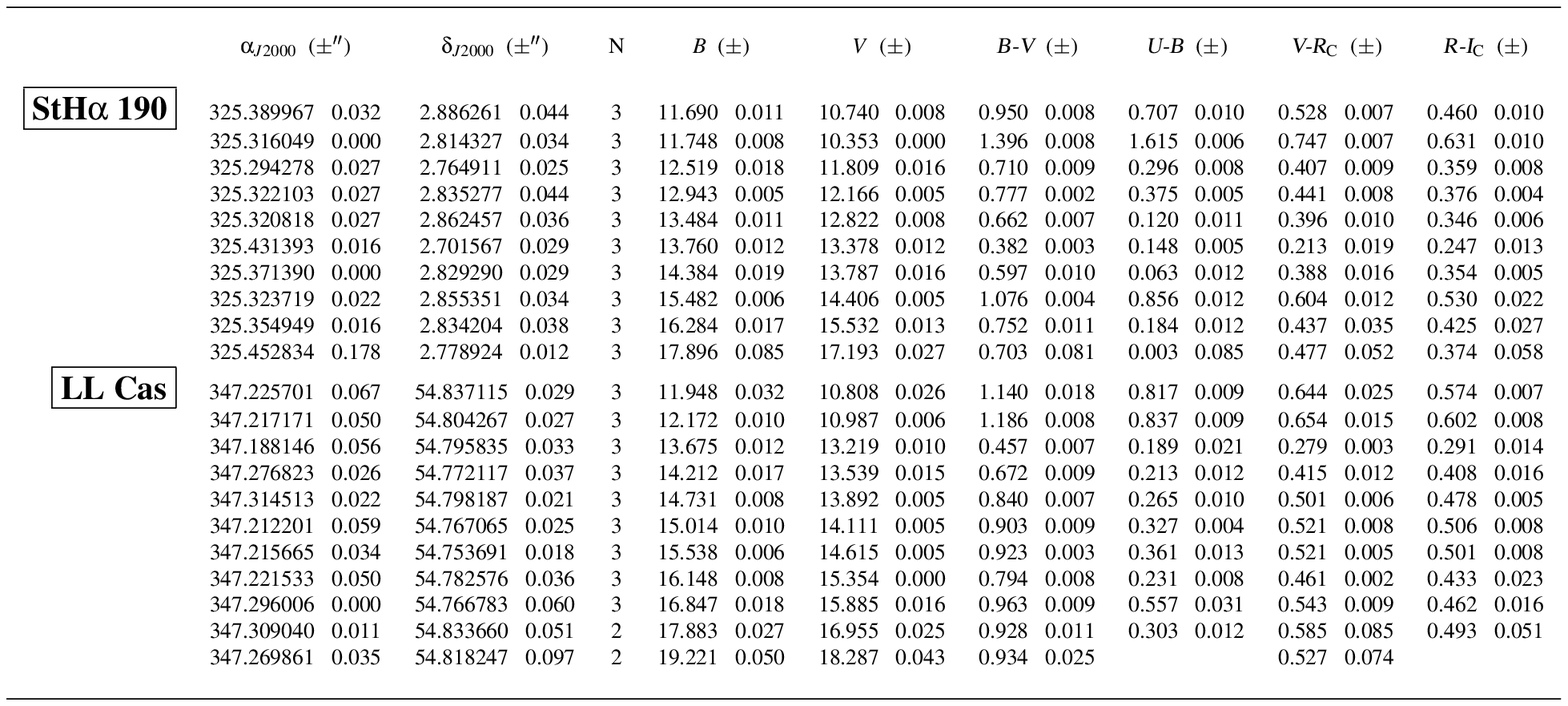,width=18cm}}
\end{table*}

\begin{table*}
\caption[]{Selected non-variable Tycho stars to supplement the sequences in
Table~2. The $B_T, V_T$ are transformed to the $B_J, V_J$ on the standard
Johnson UBV system using the relations given in the Hipparcos/Tycho
Catalogue. $U_J, R_C, I_C$ on the Johnson {\sl UBV} and Cousins {\sl RI}
systems are estimated from the Caldwell et al. (1993) transformation
functions.}
\centerline{\psfig{file=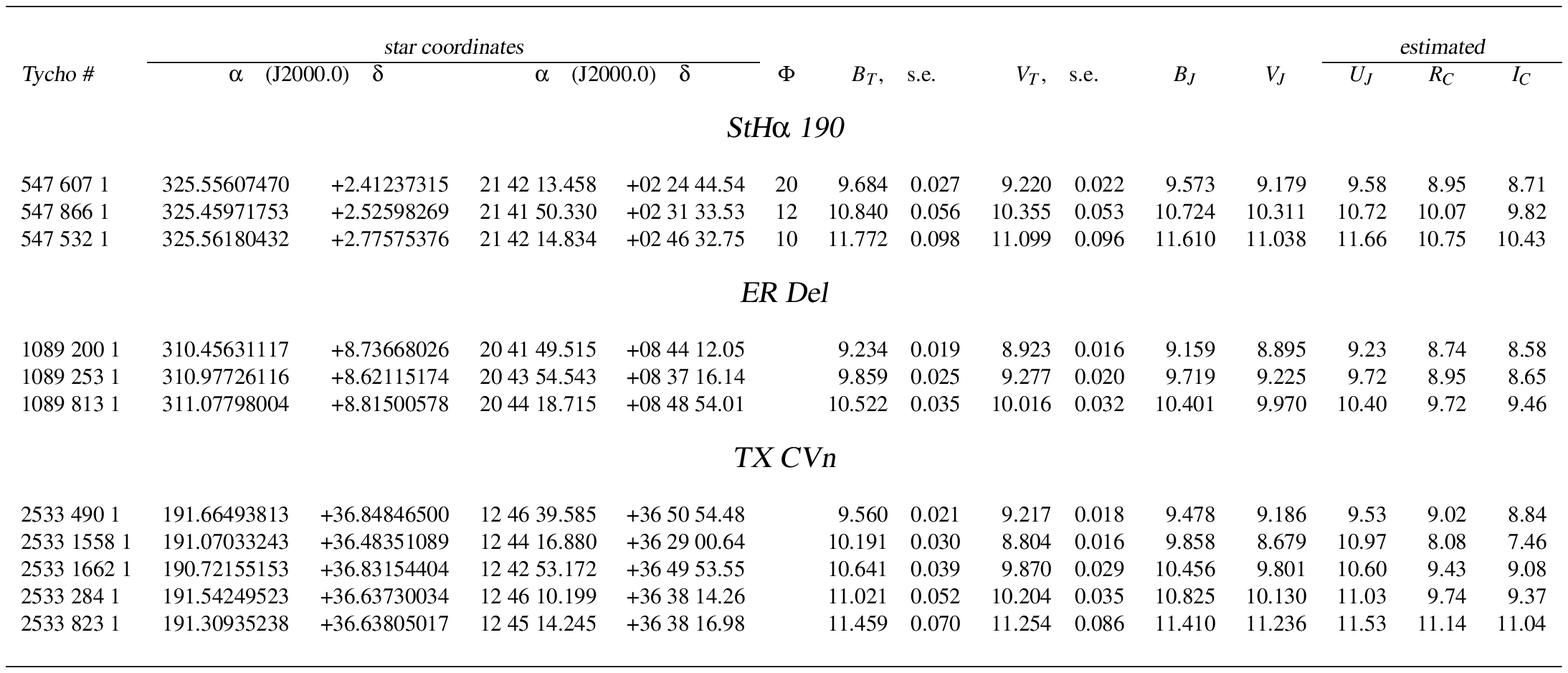,width=18cm,height=7cm}}
\end{table*}

\section{Notes on individual symbiotic stars}

A few individual notes follow on the basic photometric behavior of the
program symbiotic stars. They are intended as simple guidelines for
observers planning an observational campaign and cannot by any means 
be considered as an exhaustive review. A collected history of many symbiotic
stars and a compilation of photometric information available in literature 
has been presented by Kenyon (1983, 1986).

While calibrating the photometric comparison sequences for this paper we
have also collected data on the program symbiotic stars. These $UBV(RI)_{\rm
C}$ data will be presented and discussed elsewhere together with similar
data for more than 130 symbiotic stars observed from ESO and Asiago. From
such $UBV(RI)_{\rm C}$ survey of symbiotic stars (hereafter indicated as
MHZ: Munari, Henden and Zwitter, in preparation), we report in the following
average $V$ and $B-V$ values for the year 2000 for the reader's benefit.
Another large multi-epoch $UBV(RI)_{\rm C}$ and $JHKL$ photometric survey
has been presented by Munari et al. (1992).\\

\underline{\sl V471 Per}. Discovery and numbering of variable stars in the
Perseus constellation has not yet progressed so much to reach entry {\sl V741}~Per 
to which this variable has been erroneously confused so frequently in literature,
probably arising from a mis-print in the Allen (1984) catalogue of symbiotic 
stars.

There are not much informations available on the photometric behavior of
the system, which is usually observed around $V=13.0/13.4$ and
$B-V$=+0.90/+1.00. A moderate activity phase was observed in the 90ies,
when on 1992 the object started a slow rise reaching $V=12.4$ in 1994-95 and
declined to $V=13.3/13.4$ by 1996. The orbital period is unknown.  
MHZ lists $V=13.10$ and $B-V$=+1.00.

\underline{\sl GH Gem}. Proposed by Kenyon (1986) as a possible symbiotic
star, it has been observed by amateurs since then. VSNET and VSOLJ databases
show the star stable at $V\sim$12.4 from 1986 to 1990, when the star entered
a gradual brightness decrease to $V\sim$13.8 (reached by late 1993) and
recovering to $V\sim$12.9 by 1995. Descending slowly to $V\sim$13.2 by 1999,
in early 2000 it rapidly dropped to $V\sim$14.0. Not much else is known
about the photometric behavior of this star. MHZ reports $V=14.01$ and
$B-V$=+1.00. A spectrum for GH~Gem from the 3200-9100 spectral atlas of
symbiotic stars of Munari and Zwitter (2001) show a continuum of a K-type
giant with a weak H$\alpha$ in emission.

\underline{\sl BX Mon}. Its orbital period is around 1400 days (Mayall 1940,
Dumm et al. 1998). The lightcurve is varied, with relatively quite phases
($V\sim$12) interspersed with periods of rapid and large variability
(13.0$\leq V \leq$10.2). A fast rise, large amplitude event occurred in
1999, when BX~Mon rose to $V=9.9$ in a pattern resembling an outburst. The
photometric behavior of this bright object remain poorly known and
understood. It could be an eclipsing object according to Kenyon (1986).  
MHZ reports $B=12.36$ and $B-V$=+1.13.

\underline{\sl V694 Mon}. The object attracted much attention when hugely
blushifted absorption components (up to --6000 km sec$^{-1}$, Tomov et al.
1990) were discovered to flank the emission line spectrum during the 1990
bright phase, when the star rose to $B\sim 9.5$ mag.  Doroshenko et al.
(1993) have discovered a $\bigtriangleup B \sim$ 0.8 mag sinusoidal
variation that they attribute to manifestation of a reflection effect
following the ephemeris:
\begin{equation}
T(min) = 2437455 \ + \ 1930 \times E
\end{equation}
The photometric behavior of V694 Mon over the last century has been quite
active, spanning a range of $\sim$4 mag. The whole photometric history is
covered by Luthardt (1991), Doroshenko et al. (1993) and Tomov et al.
(1996). V694~Mon is well known for its rapid spectral variability, jet
ejection and quasi-periodic flickering behavior (e.g. Michalitsianos et al.
1993, Tomov et al. 1995). MHZ lists $V=10.81$ and $B-V$=+0.45.

\underline{\sl TX CVn}. This bright, isolated object has a relatively flat
quiescence at $V\sim$10. Moderate amplitude outbursts have been reported for
1920 (7 years to return to quiescence), 1945 (4 years duration), 1952, 1962
and 1986 (4 years rising, 5 years descent). At maximum the object may reach
$V\sim$8 mag. MHZ reports $V=10.10$ and $B-V=$+0.69 (it is worth noticing that
such a moderate $B-V$ color suggests that the zero point of the $m_{pg}$
magnitude scale adopted by Mumford (1956) could be too faint by 0.7 mag
compared to modern $B$ values). Apart from the frequent outbursts, not much
is known about the other photometric properties of TX CVn. Kenyon and Garcia
(1989) have spectroscopically determined an orbital period of 199 days, the
shortest known among symbiotic stars.

\underline{\sl IV Vir}. The orbitally modulated lightcurve (P=282 days) of this bright 
symbiotic star ($V\sim10.8$) is quite interesting: at longer
wavelengths it is dominated by the ellipsoidal distortion of the cool giant
filling its Roche lobe ($\bigtriangleup y =$0.15 mag, with the
classical pattern of two-maxima/two-minima per period; Niehues et al. 1992),
while at shorter wavelengths the reflection effect takes over
($\bigtriangleup u =$1.6 mag, with a sinusoidal pattern; Smith et al. 1997).
The ephemeris for maxima in $u$ is
\begin{equation}
T(max) = 2449158 \ + \ 282 \times E
\end{equation}
The historical lightcurve and occurrence of past outbursts are unknown.
MHZ lists $V=10.71$ and $B-V=$+1.41.

\underline{\sl UKS Ce1}. The photometric properties and history of this carbon
symbiotic stars are unknown. MHZ reports $V=15.89$ and $B-V=$+1.88.
A field star of similar brightness lies 5.9 arcsec from UKS~Ce1 toward SSW,
at $\alpha=$243.871767 ($\pm$0.055 arcsec) and $\delta=-$22.205167
($\pm$0.177 arcsec), with magnitudes $V=$16.732 ($\pm$0.023),
$B-V=+$0.840 ($\pm$0.033), $U-B=+$0.286 ($\pm$0.045), $V-R=+$0.606 ($\pm$0.030)
and $R-I=+$0.482 ($\pm$0.042).

\underline{\sl V503 Her}. In the GCVS it is reported as a long period
variable varying between 14.4$\leq B \leq$13.1. Kenyon (1986) suggested its
possible symbiotic nature but Bond (1978) and Munari and Zwitter (2001)
observed an early M spectrum without emission lines, so its nature remains
unclear. The VSOLJ data for 1988-1990 shows a relatively constant brightness
at $V=12.3$, with a slightly brighter phase at $V=11.7$ in 1987 and a few
doubtful reports about very faint states. The more recent VSNET lightcurve
reports a $V\geq 14$ in 1997 and 1999, $V\sim 12.3$ in 1998 and $V\sim 12.7$
in 2000. MHZ lists $V=12.67$ and $B-V=$+1.28.

\underline{\sl UU Her}. After the initial discovery by Reinmuth (1926)
reporting the object to vary between 16.0$\leq m_{pg} \leq$14.6, not much
else has appeared in literature.  The photometric properties of this
symbiotic star remain essentially unknown. MHZ gives $V=15.47$ and
$B-V=$+1.79.

\underline{\sl DT Ser}. This is an interesting case. The GCVS lists the
variability range as 13.9$\leq m_{pg} \leq$13.2. Cieslinski et al. (1997)
have shown that the symbiotic star is indeed a faint star ($B=$15.91,
$V=$15.40) about 5 arcsec from a much brighter field star (that we have
measured at $B=$13.550$\pm$0.003, $V=$12.772$\pm$0.014,
$U-B=$+0.109$\pm$0.004, $V-R=$+0.462$\pm$0.017 and $R-I=$0.475$\pm$0.006).
The $B=13.55$ of the nearby field star (not variable between the three
observations we have obtained in separate nights) well matches the mean
brightness of DT~Ser listed in the GCVS. Thus, we believe that GCVS data
refer to the combined light of ($a$) the much brighter, probably not
variable field star, {\sl and} ($b$) the faint, nearby symbiotic star that
cannot be easily resolved under normal seeing condition and with small
telescopes. If the nearby field star is not itself variable, than DT~Ser
must vary by a large amount to account for the observed $\bigtriangleup
m_{pg}=$0.7 mag, as if undergoing outbursts, and indeed the slow, long term
variability exhibited by the pair combined light in the VSOLJ data
(up-and-down time of $\sim$8 years) could support this interpretation. MHZ
lists $V=$16.23 and $B-V=$+0.37.

\underline{\sl YY Her}. The complete photometric history of YY~Her over
1890-1996 has been reconstructed by Munari et al. 

\begin{figure*}[!h]
\centerline{\psfig{file=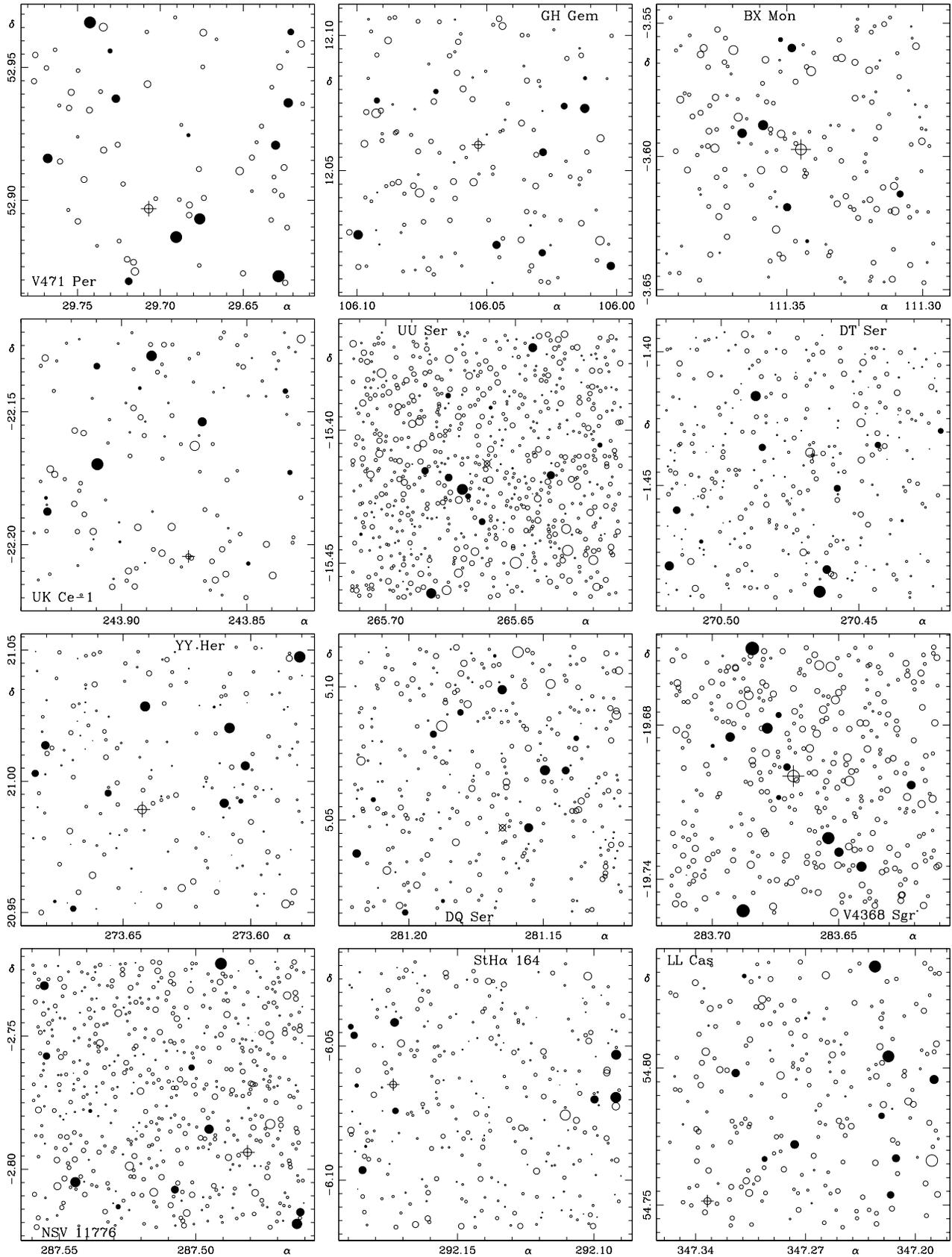,width=17cm}}
\caption[]{Finding charts for the {\sl UBV(RI)$_{\rm C}$} comparison
photometric sequence around the program symbiotic stars. The fields are
in the same order as in Table~1. North is up and East to the left,
with an imaged field of view of 6.0$\times$6.0 arcmin and
a 6.6$\times$6.6 coordinate grid.
Stars are plotted as open circles of diameter proportional to the
brightness in the $V$ band. The stars making up the photometric sequence
(see Table~2) are plotted as filled circles.}
\end{figure*}

\begin{figure*}[!h]
\centerline{\psfig{file=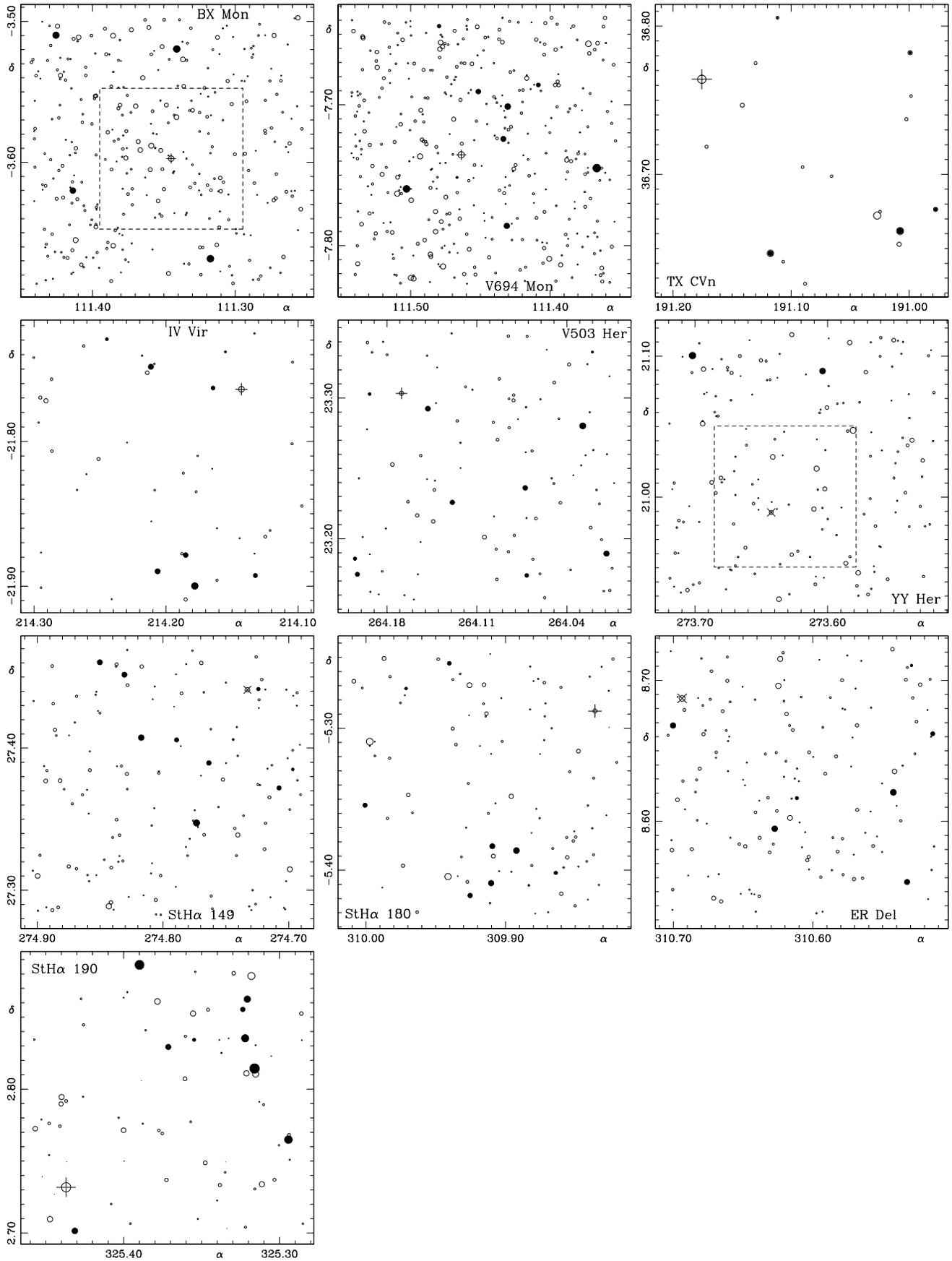,width=17.5cm}}
\caption[]{Several of program symbiotic stars reach maximum magnitudes
which impose looking for suitably bright comparison stars over a wider
field. The symbols are the same as in Figure~1, with
an imaged field of view of about 11.4$\times$11.4 arcmin and a
12$\times$12 arcmin coordinate grid. The zoomed-in 6.0$\times$6.0 arcmin
for BX~Mon and YY~Her presented in Figure~1 is outlined by a dashed
square.}
\end{figure*}

\clearpage

\noindent
(1997). The quiescent
lightcurve (at $V=13.2$) is modulated by a $\bigtriangleup V=$0.3 sinusoidal
reflection effect, with minima given by the ephemeris
\begin{equation}
T(min) = 2448945(\pm 10) \ + \ 590(\pm 3) \times E
\end{equation}
YY~Her underwent outbursts in 1914-19, 1930-33, 1981-82 and 1993-96
with additional bright phases in 1890, 1903, 1942, 1954, 1965 and 1974.
Each outburst has its own lightcurve, different from the others. A few
enigmatic brightness drops brought the star down to $V=14.8$.
MHZ gives $V=$12.93 and $B-V=$+1.38.

\underline{\sl StH$\alpha$ 149}. In spite of its brightness (MHZ lists
$V=$12.06 and $B-V$=+1.55), it is photometrically unknown.

\underline{\sl DQ Ser}. This moderately bright symbiotic star has been
so far largely ignored by observers and its photometric properties
are unknown. The GCVS gives a variability range 16.0$\leq m_{pg} \leq$13.9.
Cieslinski et al. (1997) list $V=$14.5 and $B-V$=+1.23, while MHZ reports
$V=$14.94 and $B-V$=+1.34. 

\underline{\sl V4368 Sgr}. This probable symbiotic nova was discovered 
at $V=10.7$ in 1994 by M.Wakuda. The progenitor is not visible in ESO/SERC 
photographic atlas (limiting magnitude $\sim$21.5), and searches in the Harvard 
and Sonnerberg plate archives have failed to reveal anything at this position
since the earliest
images obtained in 1888 (Grebel et al. 1994, Hazen 1994). Since then the
object has remained close to maximum brightness, in a photometric and
spectroscopic pattern reminiscent of PU~Vul, another symbiotic nova (cf.
Grebel et al. 1994, Bragaglia et al. 1995). According to the VSOLJ database
V4368~Sgr has slowly risen to $V=10.2$ by 1997 and descent to $V=10.5$ by
1999. MHZ reports $V=10.57$ and $B-V=$+0.63.

\underline{\sl NSV 11776}. Scanty photometric information exist for this
relatively bright symbiotic star. Cieslinski et al. (1994) reported about
{\sl UBV(RI)$_{\rm C}$} photometry secured in 1991-1992 that did not
revealed variability or presence of flickering activity. MHZ lists $V=13.47$
and $B-V=$+0.92, which are identical to the mean values $V=$13.46
$B-V=$+0.92 of Cieslinski et al. for 1991-92, which suggests limited
variability for NSV~11776. Such a tight photometric stability is surprising
in view of the very intense, high ionization emission lines indicating the
presence of a very hot and luminous accreting white dwarf in the system.

\underline{\sl StH$\alpha$ 164}. The photometric history, type of variability,
presence of outbursts and orbital period are unknown. MHZ reports $B$=14.48
and $B-V$=+2.03.

\underline{\sl StH$\alpha$ 180}. Another example of a bright symbiotic star
completely unknown in its photometric properties. MHZ lists $V=$12.68 and 
$B-V=$+1.40.

\underline{\sl ER Del}. ER Del is one of the rare symbiotics whose cool giant
is an S star. It attracted attention when IUE spectra showed a high
ionization emission line spectrum, revealing the symbiotic nature (as much
as it happens with EG And). Little is known about its photometric
properties. The poorly sampled VSNET lightcurve over the last six years may
be described as a slow, linear descent from $V\sim$10.3 to $V\sim$10.8, with
fluctuations overimposed. MHZ reports $V=10.33$ and $B-V=$+1.78.

\underline{\sl StH$\alpha$ 190}. The Hipparcos/Tycho did not detect
variability for this object which is however close to the sensitivity limit,
so a limited variability (some tenths of a magnitude) can surely be
accommodated by the noise in the data.  The Tycho $B_T=11.1$ and $V_T=10.30$
when transformed to the standard UBV system ($V=10.23$, $B-V$=+0.68) are
close to the values reported by MHZ, namely $V=10.50$ and $B-V$=+0.84.
Stephenson (1986) discovered StH$\alpha$~190 at $V\sim 10.5$, Robertson \&
Jordan (1989) re-discovered it (as RJH$\alpha$~120) at $V\sim 10.1$ and
Downes \& Keyes (1988) measured it at $V=10.5$. Thus over the last 15 years
the system mean brightness has remained fairly constant, and similarly went
for the optical spectra very similar one to the other (Downes \& Keyes 1988,
Whitelock et al. 1995, Munari and Zwitter 2001), with no report of
outbursts. Variability has been instead firmly established in the infrared
by Whitelock et al. (1995) who reported $\bigtriangleup K =$0.16.
Munari et al. (2001, to be submitted) have recently discovered a fast
evolving and complex mixture of spectral signatures produced by high
variable bipolar jets and P-Cyg profiles, with indication of a high orbital
inclination and a possible 171 day orbital period. Therefore, searches for flickering 
and eclipses in the bluest photometric bands could pay dividends over a single 
observing season.
 
\underline{\sl LL Cas}. Originally listed among the planetary nebulae
it has been later reclassified as an Me star by Sabbadin et al. (1987)
and finally as a symbiotic star by Kondrateva (1992) that reports about
a $\bigtriangleup m \sim$3.5 mag variability with a period P=286.6 days.

\clearpage
\end{document}